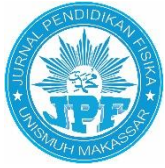

# Jurnal Pendidikan Fisika

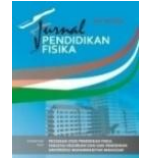

# Analyzing Students' Critical Thinking as a Basis for Developing Interactive Physics Multimedia with Generative Learning and Cognitive Conflict Strategies

## Serli Ahzari[1)], Akmam Akmam[2)*]

*Department of Physics, Universitas Negeri Padang, Padang, 25173, Indonesia*

*\*Corresponding author: akmam_db@fmipa.unp.ac.id*

*Abstract – The increasing complexity of abstract concepts in physics education and the low level of students' critical thinking skills demand innovative instructional strategies aligned with 21st-century competencies. This study aims to analyze students' critical thinking skills as the foundation for developing physics interactive multimedia using a generative learning model integrated with a cognitive conflict strategy. The research was conducted in three public high schools in Lima Puluh Kota Regency using a quantitative descriptive survey method, involving 125 eleventh-grade students. Data were collected using five validated instruments: teacher questionnaires on instructional practices and critical thinking, student learning style and attitude surveys, a multimedia needs assessment, and a critical thinking skills test. Results show that current instruction relies heavily on static and non-interactive media such as PowerPoint and videos, failing to support students' varied learning styles especially the dominant visual preference (53.06%). While students exhibit moderately positive attitudes toward physics learning (65.69%), their critical thinking skills are critically low (average score 27.80%), particularly in inference (24.00%) and evaluation (28.00%) indicators. These findings underscore the urgent need to develop visual and interactive multimedia that facilitates cognitive conflict and deeper reflection. The novelty of this study lies in linking multimedia design with cognitive strategies tailored to enhance critical thinking in physics. The study contributes to physics education by offering empirical evidence for the pedagogical integration of generative learning and cognitive conflict approaches in digital environments, paving the way for targeted multimedia interventions that promote critical reasoning in conceptually demanding physics topics.*

*Keywords: cognitive conflict; critical thinking; generative learning; interactive multimedia; physics*

## I. INTRODUCTION

Technology integration in education has become imperative in the 21st century, particularly in physics education, where abstract concepts and complex mathematical representations present unique pedagogical challenges. This integration is essential for preparing students to meet future challenges in an increasingly digital world (Elihami & Saharuddin, 2018). The importance of this research lies in addressing the critical gap between technological advancements and their practical application in physics education, where the development of student's critical thinking skills remains a primary concern for educators and policymakers alike.

Physics education faces distinctive challenges that are not encountered to the same degree in other disciplines. Students often perceive physics as highly abstract, mathematically demanding, and disconnected from everyday experiences, resulting in decreased motivation and difficulty understanding foundational concepts (Dewi et al., 2017; Maunino et al., 2023). The multidimensional nature of physics requiring simultaneous mastery of conceptual understanding, mathematical formalism, and experimental reasoning places significant cognitive demands on students. Conventional teaching approaches frequently emphasize theoretical content over practical applications, neglecting the development of critical problem-solving abilities that are essential for physics mastery (Azizah et al., 2017). Additionally, students struggle to connect theoretical concepts with real-world applications, making learning experiences less meaningful and more challenging to internalize (Wulandari et al., 2023).

Observations at several high schools in Lima Puluh Kota Regency reveal specific issues in physics education. Teachers predominantly use conventional teaching methods and simple media such as PowerPoint presentations and occasional videos, which fail to address students' diverse learning styles. These traditional approaches do not adequately visualize abstract physics concepts or engage students in active learning experiences. A preliminary assessment indicates that students' critical thinking skills remain underdeveloped, with most students struggling to independently analyze, evaluate, and construct physics knowledge. This situation calls for innovative approaches to enhance conceptual understanding and critical thinking development in physics.

Recent research has explored various solutions to address these challenges. Interactive multimedia has emerged as a powerful tool for enhancing student engagement and conceptual understanding in physics education. Studies by Nurhalimah & Rizal (2024) and Tanti et al. (2024) demonstrate that well-designed multimedia can significantly improve student engagement, conceptual understanding, and information retention in physics learning. The visual and interactive characteristics of multimedia make abstract concepts more concrete and manageable for students to comprehend. Meanwhile, pedagogical approaches such as generative learning models and cognitive conflict strategies have shown promise in developing students' higher-order thinking skills. Research by Agustin & Akmam (2024) and Akmam et al. (2023, 2024) confirm the effectiveness of these approaches in various educational contexts, particularly in science learning, where they actively engage students in knowledge construction rather than passive information reception.

Despite these advancements, significant research gaps remain in understanding how to effectively integrate generative learning models and cognitive conflict strategies within interactive multimedia environments, specifically for physics education. Previous studies by Abdul et al. (2016) have demonstrated generally positive outcomes when combining multimedia learning with active pedagogical approaches, showing significantly better learning outcomes compared to conventional methods. However, these studies have not sufficiently investigated how cognitive conflict strategies might be systematically incorporated into interactive multimedia designs to address specific misconceptions in physics. Moreover, existing research has not adequately examined how this integration might enhance critical thinking skills development in physics education contexts. The unique challenges of representing complex physics phenomena in interactive multimedia formats balancing scientific accuracy with user-friendliness and maintaining high student engagement (Buelow et al., 2018; Unal & Cakir, 2021) further complicate this integration and require specialized research attention.

This study aims to analyze students' critical thinking skills as a foundation for developing physics interactive multimedia based on a generative learning model with a cognitive conflict strategy. The research specifically addresses the question: What is the current level of critical thinking skills among high school students in Lima Puluh Kota Regency, and how can this understanding inform the development of practical physics interactive multimedia that integrates generative learning models and cognitive conflict strategies? The findings from this research are expected to contribute significantly to a broader understanding of how technology can support the development of students' critical thinking skills. Furthermore, this research is anticipated to bridge existing gaps in current educational practices and literature, serving as a reference for educators and multimedia developers in designing and implementing effective technology-based learning in physics education.

## II. METHODS

This study employs a descriptive survey method with a quantitative approach to systematically analyze students' critical thinking skills in physics learning. The research examines multiple dimensions, including learning styles, student attitudes, interactive multimedia needs, and learning model implementation as a foundation for developing physics interactive multimedia based on a generative learning model with a cognitive conflict strategy. The research procedure followed three main phases: preparation, implementation, and final analysis, as illustrated in Figure 1.

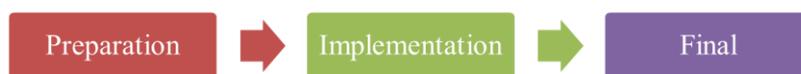

**Figure 1.** Research procedure

During the preparation phase, preliminary observations were conducted at the selected schools to identify existing teaching practices and student learning environments. The researchers conducted a comprehensive literature review on critical thinking skills, generative learning models, cognitive conflict strategies, and interactive multimedia development in physics education. Research instruments were systematically developed and validated by three physics education experts before obtaining the necessary research permits from the regional education authorities of Lima Puluh Kota Regency.

In the implementation phase, data were collected over approximately four weeks across three public high schools in Lima Puluh Kota Regency. These schools were selected based on their 2024 accreditation levels (high, medium, and low) to ensure representation of diverse educational environments. The research subjects included 125 eleventh-grade students (Phase F) enrolled in physics classes across the three schools, as presented in Table 1.

**Table 1**. Distribution of research subjects

| School | Accreditation level | Number of students |
|--------|--------------------|--------------------|
| SHS A | High | 49 |
| SHS B | Medium | 48 |
| SHS C | Low | 28 |
| Total | | 125 |

Data collection was performed using five validated instruments. First, a questionnaire was used to assess physics teachers' responses toward the Generative Learning Model with Cognitive Conflict Strategy and Critical Thinking Skills in Grade XI Senior High School Physics Learning. The questionnaire consisted of two main components: (1) a section on the Generative Learning Model with Cognitive Conflict Strategy in Physics Learning with 6 statement points, and (2) a section on Critical Thinking Skills in Physics Learning with 5 statement points. This instrument utilized a 5-point Likert scale.

Second, a questionnaire was used to analyze the need for interactive multimedia in Grade XI Phase F Senior High School Physics Learning, which included 18 statements with multiple-choice answers and short responses. Third, a questionnaire was employed to analyze students' attitudes toward physics learning, adopted from the CLASS (Colorado Learning Attitude about Science Survey), using a 5-point Likert scale (1 = Strongly Disagree, 2 = Disagree, 3 = Neutral, 4 = Agree, 5 = Strongly Agree). This questionnaire contained 30 statements, with one distractor statement.

Fourth, a learning style questionnaire was administered to Grade XI Phase F Senior High School students (adopted from the Student Learning Style Questionnaire by Akhmad Sugianto, S.Pd., M.Pd), which consisted of 14 statements with three answer choices for each statement. Fifth, a critical thinking skills test sheet was used, consisting of 5 contextual physics phenomenon-based questions in narrative form.

The final phase involved comprehensive data analysis. The collected data were analyzed descriptively to obtain a comprehensive statistical overview of students' critical thinking skills. For critical thinking skills analysis, initial scores were converted to a 100-point scale using the following formula by Purwanto (2008):

$$S = \frac{R}{N} x \ 100\%$$

Where:
S = Percentage of critical thinking skills
R = Total score obtained by the student
N = Maximum test score

The calculated percentages were then categorized according to the critical thinking skills assessment criteria, as shown in Table 2.

**Table 2**. Critical thinking skills assessment criteria

| Interval | Criteria |
|----------|----------|
| 86% - 100% | Very high |
| 76% - 86% | High |
| 60% - 75% | Medium |
| 55% - 59% | Low |
| ≤54% | Very low |

(Purwanto, 2008)

## III. RESULTS AND DISCUSSION

Based on the questionnaire analysis administered to three physics teachers teaching 11[th] Grade Phase F in three different high schools in Lima Puluh Kota Regency, it was found that teachers primarily use simple media such as PowerPoint presentations, learning videos, audio, and real objects. The use of such media in physics learning tends not to meet students' diverse learning styles and does not optimally support the development of critical thinking skills. PowerPoint presentations are often static and function only as one-way tools, making students passive listeners without active interaction. Although learning videos can be more engaging, their content is linear and offers limited opportunities for students to interact or explore the material independently. Audio usage is restricted because it relies solely on hearing, making it less suitable for students with visual or kinesthetic learning styles. Real objects, while more concrete, are often limited in accessibility and challenging to manage effectively in the classroom.

Field observations revealed that teachers have not widely adopted interactive multimedia due to time constraints for learning and developing new media, compounded by a lack of technological facilities in schools and teachers limited technical skills. Without adequate training and facilities, teachers are more likely to use simple learning media they have already mastered. The lack of engagement inhibits interactivity in learning, which is crucial for promoting students' critical thinking skills (Ennis, 2011). Therefore, the development of interactive multimedia, such as virtual labs, simulations, and interactive tutorials, based on generative learning models and cognitive conflict strategies, is urgently needed to facilitate student engagement and support the development of students' critical thinking skills.

Regarding the implementation of learning models, teachers frequently use discovery, problem-based, and project-based learning approaches. While these models effectively promote independent learning, they often lack depth in stimulating students' critical thinking skills. Critical thinking skills are essential in physics learning, which requires students to analyze, evaluate, and solve problems with a more structured and in-depth approach (Fisher, 2015). The weakness of existing learning models is the lack of cognitive challenges that can prompt students to think critically. Therefore, developing interactive multimedia based on generative learning models with cognitive conflict strategies is essential, as this strategy can create cognitive conflicts that encourage students to think more deeply, explore new ideas, and discover innovative solutions (Mayer, 2020).

Interactive multimedia enriches students' learning experiences by providing an active and participatory learning environment, crucial for enhancing critical thinking skills. Specifically, interactive multimedia fosters critical thinking through several mechanisms: (1) scenario-based simulations that allow students to analyze complex situations and make informed decisions, (2) interactive data visualization features that help students identify patterns and cause-effect relationships, (3) real-time feedback systems that encourage students to evaluate their solutions and identify conceptual misconceptions, and (4) multimedia-based collaborative tasks that require students to communicate their reasoning and respond to alternative perspectives (Ahzari & Asrizal, 2023; Asrizal et al., 2025; Hamdani et al., 2022; Rogti, 2024). The ability of interactive multimedia to integrate elements of problem-based learning and discovery learning also facilitates the application of Bloom's higher-order cognitive taxonomy analysis, evaluation, and creation at the core of critical thinking. Given the limitations of teachers' facilities and technical abilities, interactive multimedia can be an important tool to help them adopt this learning model more effectively.

Student characteristics analysis is necessary to understand students' traits as a basis for considerations in developing learning models. The student characteristics examined include their learning styles and attitudes toward physics learning. The respondents were 125 11th-grade Phase F students from three different high schools in Lima Puluh Kota Regency. The analysis of students' learning styles is presented in Figure 2 below.

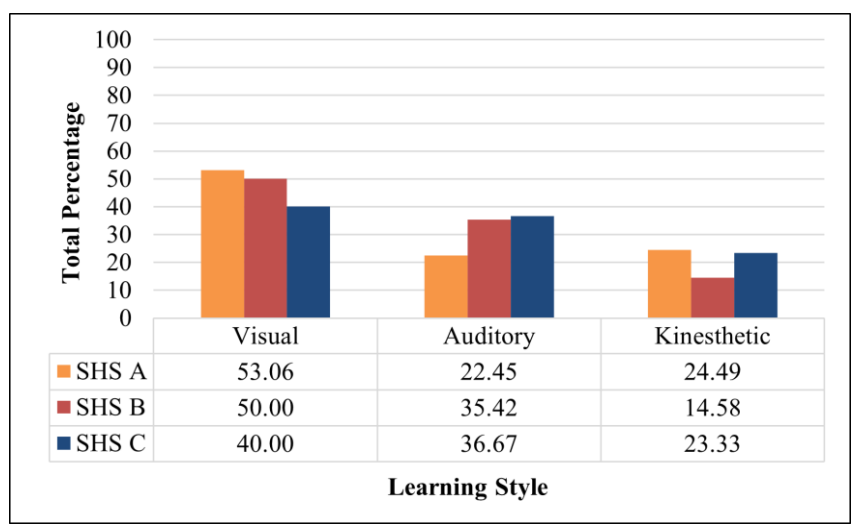

**Figure 2.** Students' learning style

The analysis reveals that students possess diverse learning styles, as evidenced by the distribution across the three schools. This diversity indicates the necessity for learning models and educational media that can accommodate various learning styles to ensure the effectiveness of the teaching and learning process. With most students exhibiting a visual learning style (53.06% at SHS A), followed by auditory and kinesthetic styles, it is crucial to develop interactive multimedia that is visually engaging and integrates auditory and kinesthetic elements. This approach will help create a more inclusive and dynamic learning experience, enhancing student engagement in the learning process and developing the essential critical thinking skills crucial in physics learning. Furthermore, regarding students' attitudes toward physics learning, based on the questionnaire distributed to students, their attitude toward physics learning is quite positive, with 65.69% showing a favorable disposition. The graph analyzing students' attitudes across different indicators is presented in Figure 3.

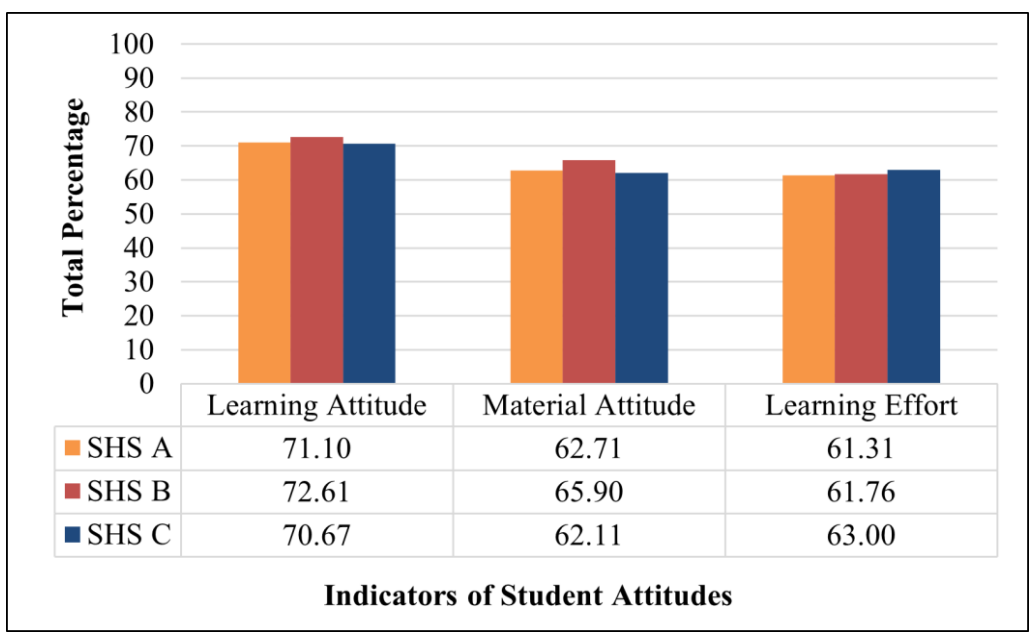

**Figure 3.** Indicators of student attitudes

Based on the questionnaire distributed to 125 11th Grade Phase F students across three schools in Lima Puluh Kota Regency, the table displays student attitude indicators in three senior high schools (SHS A, SHS B, and SHS C) in three categories: learning attitude, material attitude, and learning effort. The average student learning attitude reached 71.46%, with SHS B recording the highest percentage (72.61%), indicating a reasonably positive attitude toward the learning process.

However, in the material attitude category, the average was only 63.57%, signaling room for improvement in student engagement with the subject matter, with SHS B again holding the highest percentage (65.90%). Regarding learning effort, students averaged only 62.02%, with SHS C recording the highest percentage (63.00%), demonstrating challenges in learning motivation. Overall, while students' learning attitudes are relatively positive, there is a clear need to enhance their attitudes toward learning materials and their learning efforts.

These findings underscore the importance of developing more interactive and engaging learning strategies, such as interactive multimedia based on generative learning models, to support and improve student engagement and positive attitudes toward learning. Positive attitudes toward physics learning can increase student motivation and effort in understanding the concepts taught (Husein et al., 2017). Analyzing students' attitudes toward physics learning is crucial to understanding their readiness to engage in interactive multimedia-based learning.

The students' critical thinking skills were analyzed from test question sheets given to 11th-grade Phase F students in three different state high schools in Lima Puluh Kota Regency. The aim was to measure students' critical thinking skills. Five indicators assessed critical thinking skills:

interpretation, analysis, evaluation, inference, and explanation (Facione, 2011). Based on the test results, students' critical thinking skills remain low, with approximately 27.80% proficiency. The graph analyzing critical thinking skills by indicator is presented in Figure 4.

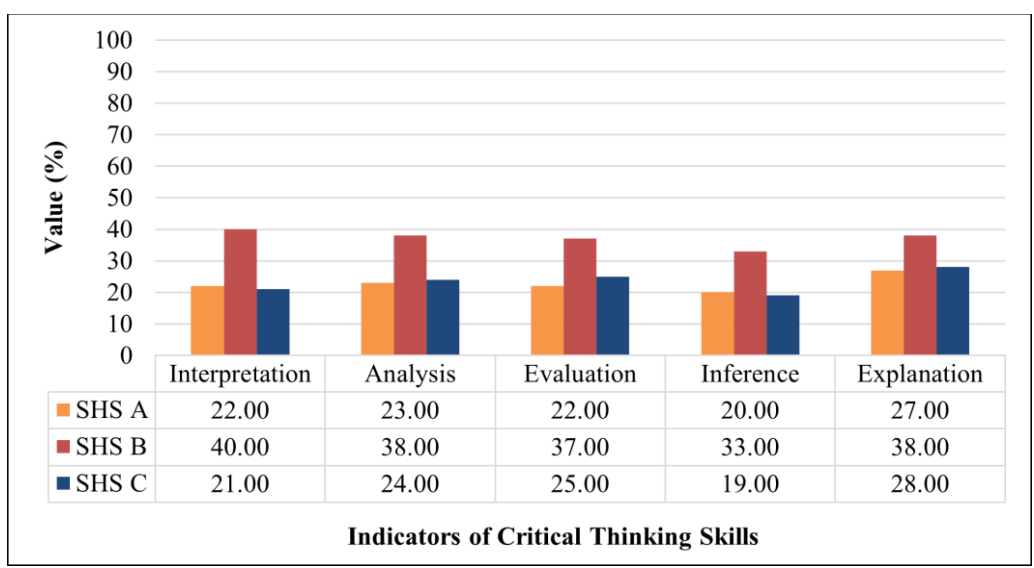

**Figure 4.** Critical thinking skills analysis results per indicator

Based on the questionnaire distributed to 125 11th Grade Phase F students across three schools in Lima Puluh Kota Regency, the critical thinking skills analysis revealed the following: First, the average scores for interpretation, analysis, and evaluation indicators were 27.67%, 28.33%, and 28.00%, respectively, all of which fall into the low category. Second, the average score for the inference indicator was 24.00%, also categorized as low. Third, the explanation indicator averaged 31.00%, which, although the highest among the five indicators, still remains in the low category.

Students' critical thinking skills must be significantly improved, as they should ideally be in the very high category. Therefore, there is a crucial need for interactive physics multimedia based on generative learning models with cognitive conflict strategies that can effectively enhance students' critical thinking skills. The analysis underscores the importance of developing innovative educational approaches that systematically improve students' ability to interpret, analyze, evaluate, draw inferences, and explain complex scientific concepts. The consistently low performance across critical thinking indicators suggests that teaching methods must be improved to foster the deep, analytical thinking required in physics education.

These findings align with research by Haris et al. (2024), which reveals that 75.50% of high school students in Makassar City demonstrate low critical thinking skills in physics problem-solving. The limited critical thinking abilities observed in our study (27.80%) are comparable to findings from Prabayanti et al. (2024), who found that 60% of students at SMAN 5 Sidrap fell into the low category for critical thinking skills when studying Newton's Laws. Further support comes from Gunawan et al. (2024), whose research with prospective science teachers identified persistent difficulties across multiple critical thinking indicators, including inference, advanced clarification, and strategy formulation. The research evidence confirms a consistent pattern of underdeveloped critical thinking skills in physics education across Indonesia's educational levels and contexts.

The predominance of visual learning styles among students (approximately 50% across schools studied) highlights a significant reliance on visual aids in learning, although specific percentages vary in the literature. This finding aligns with research by Tanti et al. (2024), which demonstrates that interactive visual elements in learning resources significantly enhance students' conceptual understanding of physics. Students with visual learning tendencies benefit substantially from instructional strategies incorporating visual representation diagrams, animations, and multimedia resources that make abstract physics concepts more concrete and accessible (Bobek & Tversky, 2016; Bouchée et al., 2022; Nasution et al., 2025). Laurenty et al. (2024) similarly found that interactive and visually engaging materials improve conceptual understanding, suggesting educators should integrate these elements into their instructional approaches. Research indicates that tailored multimedia materials in physics lessons have dramatically boosted student engagement (from 45% to 85%) and significantly raised post-test scores, demonstrating better comprehension and retention of concepts (Nasution et al., 2025).

Furthermore, this comparative analysis revealed that explanation skills (31.00%) were relatively stronger than inference skills (24.00%), consistent with findings from Agustin & Akmam (2024) and Raisal et al. (2023), who noted that students typically develop explanation abilities before mastering more complex inferential reasoning in physics education. While visual learning preferences are important, conceptual understanding in physics is also influenced by cognitive styles and student interest (Devy et al., 2022), with cooperative and interactive learning strategies further enhancing comprehension when combined with visual elements, particularly for students with converger learning styles (Kade et al., 2019). This integrated pedagogical approach clarifies complex ideas and fosters deeper engagement and more robust learning outcomes in physics education.

The moderately positive attitudes toward physics learning (65.69%) but low scores in learning effort (62.02%) reflect a disconnect between students' theoretical appreciation of physics and their willingness to engage deeply with the subject. This observation is consistent with Husein et al. (2017), who found that positive attitudes alone do not translate to improved learning outcomes without engaging learning experiences. These findings suggest that interactive multimedia must present content effectively and cultivate greater student engagement and motivation.

The findings of this study present several critical implications for the improvement of physics education. First, the results underscore an urgent need for instructional resources that intentionally cultivate students' critical thinking abilities, rather than focusing solely on content delivery. Materials should be structured to promote skills such as analysis, evaluation, and logical reasoning core components of scientific literacy. Second, the design of educational media must align with students' learning preferences, particularly the prevalent visual modality, by incorporating interactive visual elements that support cognitive engagement. Third, the application of generative learning models combined with cognitive conflict strategies emerges as a pedagogical approach with significant potential to address the low levels of critical thinking identified in this study.

The development of interactive multimedia based on these principles could offer a targeted solution to the challenges observed in current physics instruction. By integrating features that stimulate cognitive conflict, present abstract concepts through visual formats, and facilitate active student participation, such multimedia can enhance both conceptual understanding and higher-order thinking. Instructional designers and educators are therefore encouraged to adopt a learner-centered approach that embeds these strategies into technology-based learning tools. This approach offers a path toward more effective physics education, fostering not only knowledge acquisition but also the development of essential critical thinking competencies.

## IV. CONCLUSION AND SUGGESTION

This study analyzed students' critical thinking skills to inform the development of physics interactive multimedia based on a generative learning model integrated with a cognitive conflict strategy. The results revealed three key findings. First, physics learning at the observed schools remains dominated by the use of simple and static instructional media such as PowerPoint and videos, which fail to engage students actively or support the development of higher-order thinking. Second, student learning preferences are highly diverse, with visual learning styles being the most dominant (53.06%), indicating the need for multimedia that is not only content-rich but also visually engaging and interactive. Third, while students show moderately positive attitudes toward physics learning (65.69%), their critical thinking skills remain critically low (27.80%), especially in inference and evaluation indicators.

These results underscore the necessity of designing physics interactive multimedia that integrates visual, interactive, and cognitively engaging features to foster meaningful learning. However, the study is limited to a descriptive survey within a specific regional context, without testing the effectiveness of the proposed multimedia design. Additionally, the study does not explore in-depth correlations between critical thinking and other cognitive or affective variables.

Future research should focus on developing and empirically testing prototypes of interactive multimedia based on generative learning and cognitive conflict strategies, using experimental or mixed-method designs to measure their impact on students' learning outcomes. This study contributes to physics education by offering a research-based framework for instructional media development tailored to students' cognitive needs, particularly in enhancing critical thinking skills in conceptually challenging content. By aligning pedagogical design with student characteristics, the findings provide a foundation for more responsive and transformative physics instruction.

## ACKNOWLEDGMENTS

The authors gratefully acknowledge the financial support provided by the Directorate of Research, Technology, and Community Service, the Directorate General of Higher Education, Research and Technology, and the Ministry of Education, Culture, Research, and Technology according to research implementation contract number 069/E5/PG.02.00.PL/2024, and Universitas Negeri Padang (PPS-PTM 2024) number 2638/UN35.15/LT/2024.

## REFERENCES

Abdul, B., Adesope, O. O., Thiessen, D. B., & Vanwie, B. J. (2016). Comparing the effects of two active learning approaches. *International Journal of Engineering Education*, *32*(2(A)), 654–669.

Agustin, N., & Akmam, A. (2024). The influence of a generative learning model based on wave material cognitive conflict on student learning outcomes at SMAN 5 Payakumbuh. *Physics Learning and Education*, *2*(2), 81–88. https://doi.org/10.24036/ple.v2i2.127

Ahzari, S., & Asrizal, A. (2023). Developing stem-integrated interactive multimedia to improve students' data literacy and technology literacy. *Jurnal Eksakta Pendidikan (JEP)*, *7*(1), 63–73. https://doi.org/10.24036/jep/vol7-iss1/737

Akmam, A., Afrizon, R., Koto, I., Setiawan, D., Hidayat, R., & Novitra, F. (2024). Integration of conflict in generative learning model to enhancing students' creative thinking skills. *Eurasia Journal of Mathematics, Science and Technology Education*, *20*(9), 1-19. https://doi.org/10.29333/ejmste/15026

Akmam, A., Hidayat, R., Mufit, F., Anshari, R., & Jalinus, N. (2023). Effect of generative learning models based on cognitive conflict on students' creative thinking processes based on metacognitive. *Journal of Physics: Conference Series*, *2582*, 1-13. https://doi.org/10.1088/1742-6596/2582/1/012058

Asrizal, A., Annisa, A., Ahzari, S., & Helma, H. (2025). Interactive multimedia sound and light waves integrated stem to develop concept understanding and literacy skills of students. *Journal of Turkish Science Education*, *22*(1), 18–32. https://doi.org/10.36681/tused.2025.002

Azizah, R., Yulianti, L., & Latifah, E. (2017). Kemampuan pemecahan masalah melalui

pembelajaran interactive demonstration siswa kelas X SMA pada materi kalor. *Jurnal Pendidikan Fisika dan Teknologi*, *2*(2), 55–60. https://doi.org/10.29303/jpft.v2i2.289

Bobek, E., & Tversky, B. (2016). Creating visual explanations improves learning. *Cognitive Research: Principles and Implications*, *1*(1), 1-14. https://doi.org/10.1186/s41235-016-0031-6

Bouchée, T., Smits, L. de P., Thurlings, M., & Pepin, B. (2022). Towards a better understanding of conceptual difficulties in introductory quantum physics courses. *Studies in Science Education*, *58*(2), 183–202. https://doi.org/10.1080/03057267.2021.1963579

Buelow, J. R., Barry, T. A., & Rich, L. E. (2018). Supporting learning engagement with online students. *Online Learning*, *22*(4), 313–340. https://doi.org/10.24059/olj.v22i4.1384

Devy, N. K., Halim, A., Syukri, M., Yusrizal, Y., Nur, S., Khaldun, I., & Saminan, S. (2022). Analysis of understanding physics concepts in terms of students' learning styles and thinking styles. *Jurnal Penelitian Pendidikan IPA*, *8*(4), 1937-1943. https://doi.org/10.29303/jppipa.v8i4.1926

Dewi, S. M., Harjono, A., & Gunawan, G. (2017). Pengaruh model pembelajaran berbasis masalah berbantuan simulasi virtual terhadap penguasaan konsep dan kreativitas fisika siswa SMAN 2 Mataram. *Jurnal Pendidikan Fisika dan Teknologi*, *2*(3), 123–128. https://doi.org/10.29303/jpft.v2i3.302

Elihami, E., & Saharuddin, A. (2018). Peran teknologi pembelajaran islam dalam organisasi belajar. *Edumaspul: Jurnal Pendidikan*, *1*(1), 1–8. https://doi.org/10.33487/edumaspul.v1i1.34

Ennis, R. H. (2011). *The nature of critical thinking: an outline of critical thinking dispositions and abilities*. University of Illinois.

Facione, P. A. (2011). *Critical thinking : What it is and why it counts*. Insight assessment.

Fisher, A. (2015). *Critical thinking: An introduction (2nd ed.)*. Cambridge University Press.

Gunawan, K. D. H., Liliasari, L., Kaniawati, I., & Riandi, R. (2024). Status of prospective science teachers' critical and creative thinking skills in energy and its integration topics. *Jurnal Pendidikan Fisika*, *12*(3), 152–162. https://doi.org/10.26618/jpf.v12i3.11521

Hamdani, S. A., Prima, E. C., Agustin, R. R., Feranie, S., & Sugiana, A. (2022). Development of android-based interactive multimedia to enhance critical thinking skills in learning matters. *Journal of Science Learning*, *5*(1), 103–114. https://doi.org/10.17509/jsl.v5i1.33998

Haris, A., Martawijaya, M. A., Dahlan, A., Yulianti, E., & Nua, M. T. P. (2024). Analysis of critical thinking skills of high school students. *Jurnal Pendidikan Fisika*, *12*(1), 23–32. https://doi.org/10.26618/jpf.v12i1.12677

Husein, S., Herayanti, L., & Gunawan, G. (2017). Pengaruh penggunaan multimedia interaktif terhadap penguasaan konsep dan keterampilan berpikir kritis siswa pada materi suhu dan kalor. *Jurnal Pendidikan Fisika dan Teknologi*, *1*(3), 221–225. https://doi.org/10.29303/jpft.v1i3.262

Kade, A., Degeng, I. N. S., & Ali, M. N. (2019). Effect of jigsaw strategy and learning style to conceptual understanding on senior high school students. *International Journal of Emerging Technologies in Learning (IJET)*, *14*(19), 4-15. https://doi.org/10.3991/ijet.v14i19.11592

Laurenty, F., Liliawati, W., Ramalis, T. R., & Suwarna, I. R. (2024). Teaching material on metaverse for motion dynamics subject for students (motion dynamic metaverse (Md-Verse). *Eduvest-Journal of Universal Studies*, *4*(7), 6191–6197. https://doi.org/10.59188/eduvest.v4i7.1346

Maunino, P., Lantik, V., & Astiti, K. A. (2023). Penerapan model pembelajaran problem solving tutor sebaya untuk pemahaman konsep siswa materi hukum kirchhoff. *Jurnal Pendidikan dan Pembelajaran IPA Indonesia*, *13*(2), 66–76. https://doi.org/10.23887/jppii.v13i2.67807

Mayer, R. E. (2020). *How can a multimedia environment facilitate learning? In the cambridge handbook of multimedia learning (3rd ed.)*. Cambridge University Press.

Nasution, E. S., Nasution, F., Harahap, T. R., & Tambunan, E. E. (2025). Language and visual representation in physics: enhancing understanding through multimedia. *International Journal of Educational Research Excellence (IJERE)*, *4*(1), 01–09. https://doi.org/10.55299/ijere.v4i1.1226

Nurhalimah, S., & Rizal, R. (2024). Implementation of focus explore reflect apply (fera) learning model assisted crocodile physics in improving students' critical thinking skills. *JIPF (Jurnal Ilmu Pendidikan Fisika)*, *9*(2), 172-180. https://doi.org/10.26737/jipf.v9i2.4771

Prabayanti, E., Usman, U., Khaeruddin, K., & Setiawan, T. (2024). Analysis of students' critical thinking abilities in physics learning: A case study at SMAN 5 Sidrap. *Jurnal Pendidikan Fisika*, *12*(3), 141–151. https://doi.org/10.26618/jpf.v12i3.15317

Purwanto, P. (2008). *Prinsip-prinsip dan teknik evaluasi pengajaran*. PT Remaja Rosdakarya.

Raisal, A. Y., Fauziah, R. N., & Kuswanto, H. (2023). Simulation of free energy of mixing for a polymer solution using a spreadsheet for learning activities. *Jurnal Pendidikan Fisika*, *12*(2), 165–170. https://doi.org/10.24114/jpf.v12i2.52810

Rogti, M. (2024). The effect of mobile-based interactive multimedia on thinking engagement and cooperation. *International Journal of Instruction*, *17*(1), 673–696. https://doi.org/10.29333/iji.2024.17135a

Tanti, T., Deliza, D., & Hartina, S. (2024). The effectiveness of using smartphones as mobile-mini labs in improving students' beliefs in physics. *JIPF (Jurnal Ilmu Pendidikan Fisika)*, *9*(3), 387-394. https://doi.org/10.26737/jipf.v9i3.5185

Unal, E., & Cakir, H. (2021). The effect of technology-supported collaborative problem solving method on students' achievement and engagement. *Education and Information Technologies*, *26*, 4127–4150. https://doi.org/10.1007/s10639-021-10463-w

Wulandari, D., Maison, M., & Kurniawan, D. A. (2023). Identifikasi pemahaman konsep dan kemampuan berargumentasi peserta didik pada pembelajaran fisika. *Jurnal Pendidikan MIPA*, *13*(1), 93–99. https://doi.org/10.37630/jpm.v13i1.817